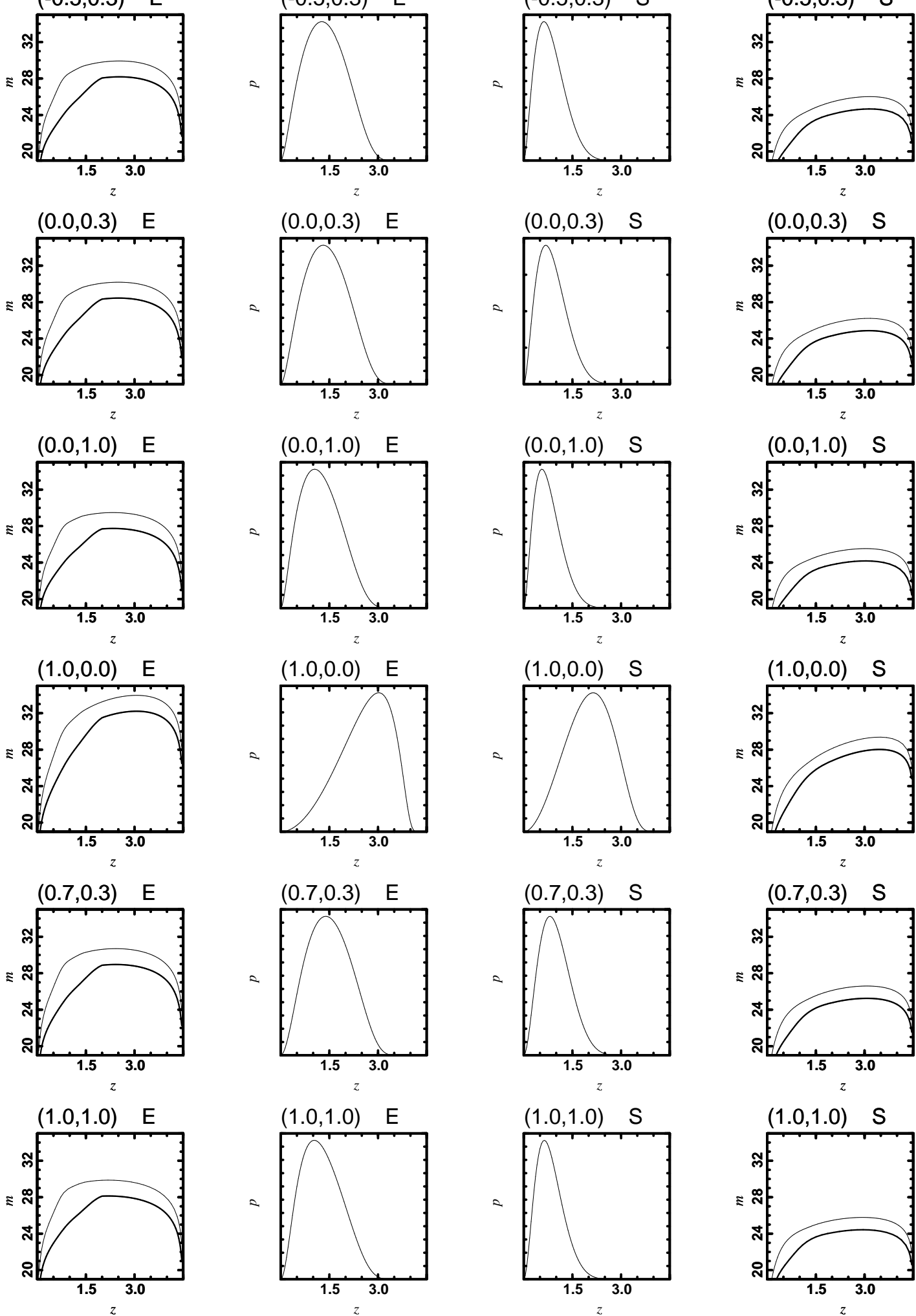

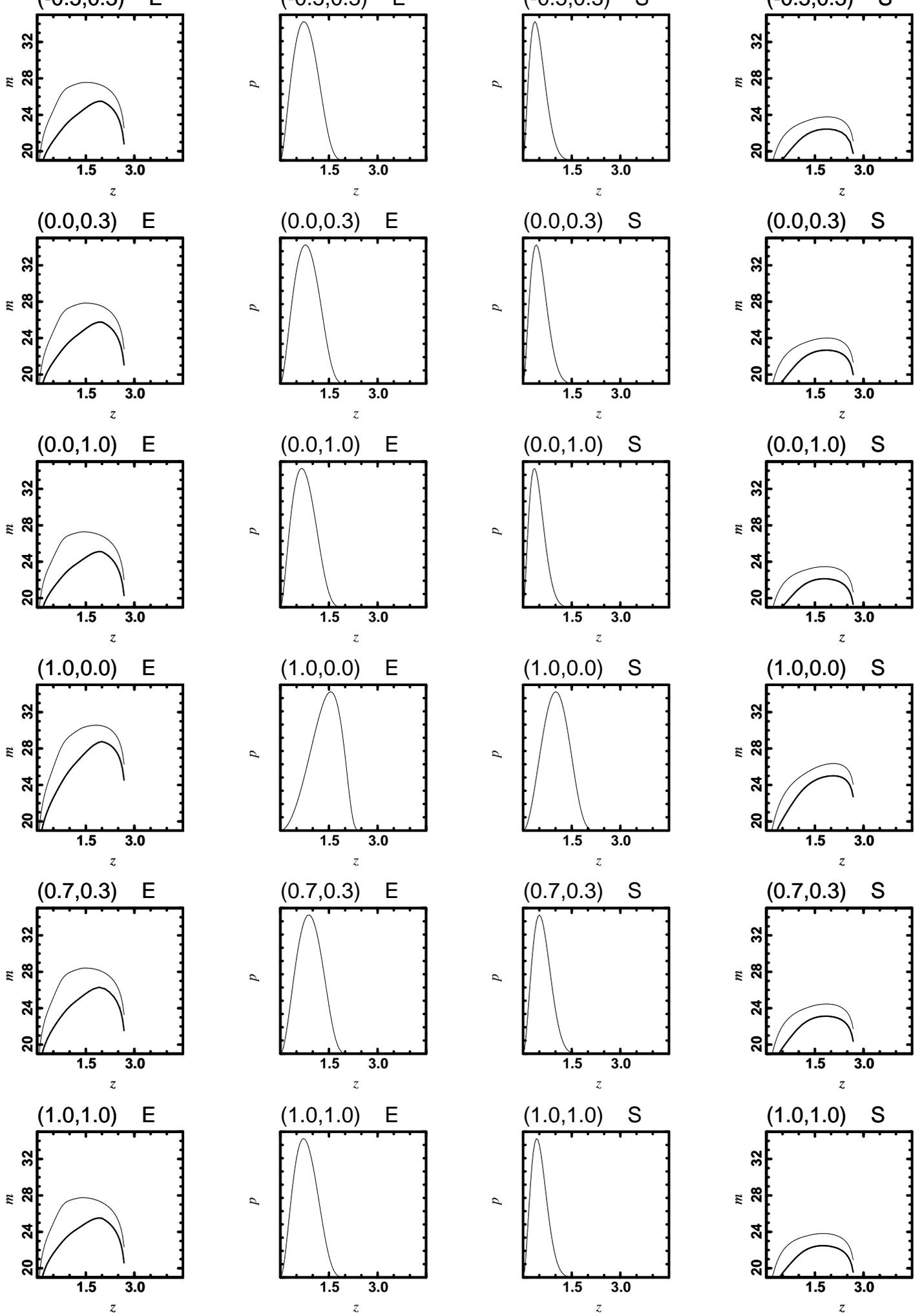

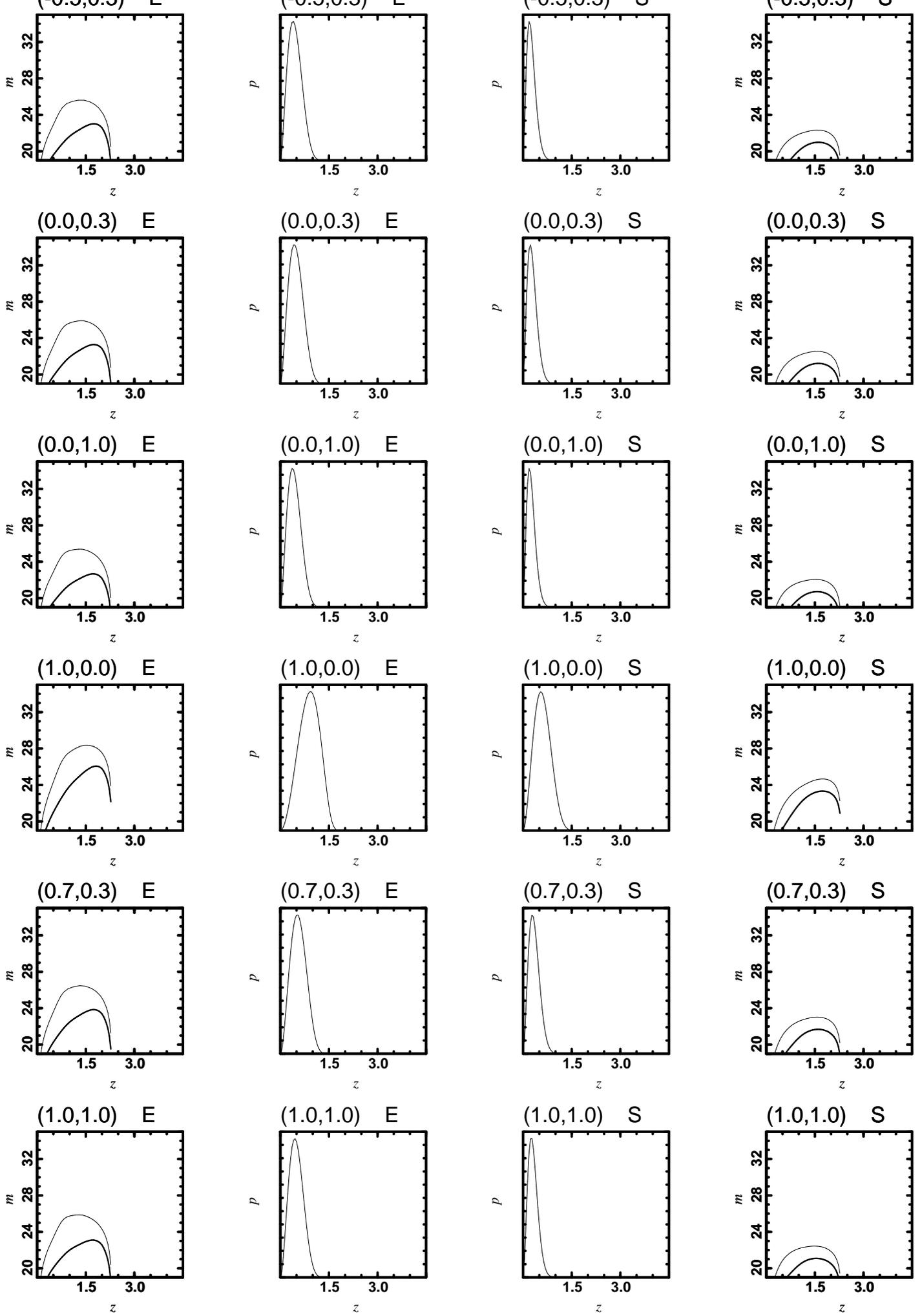

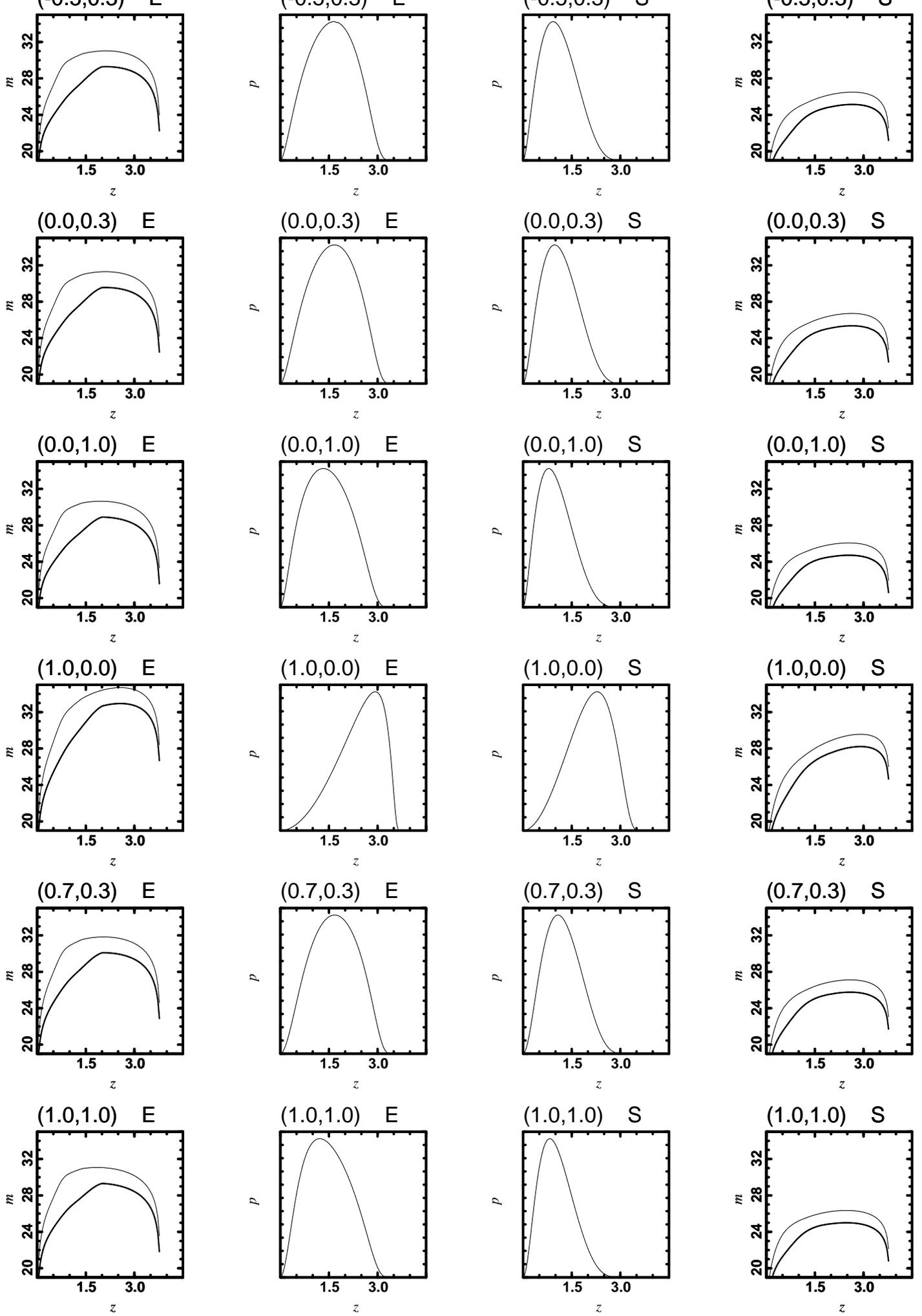

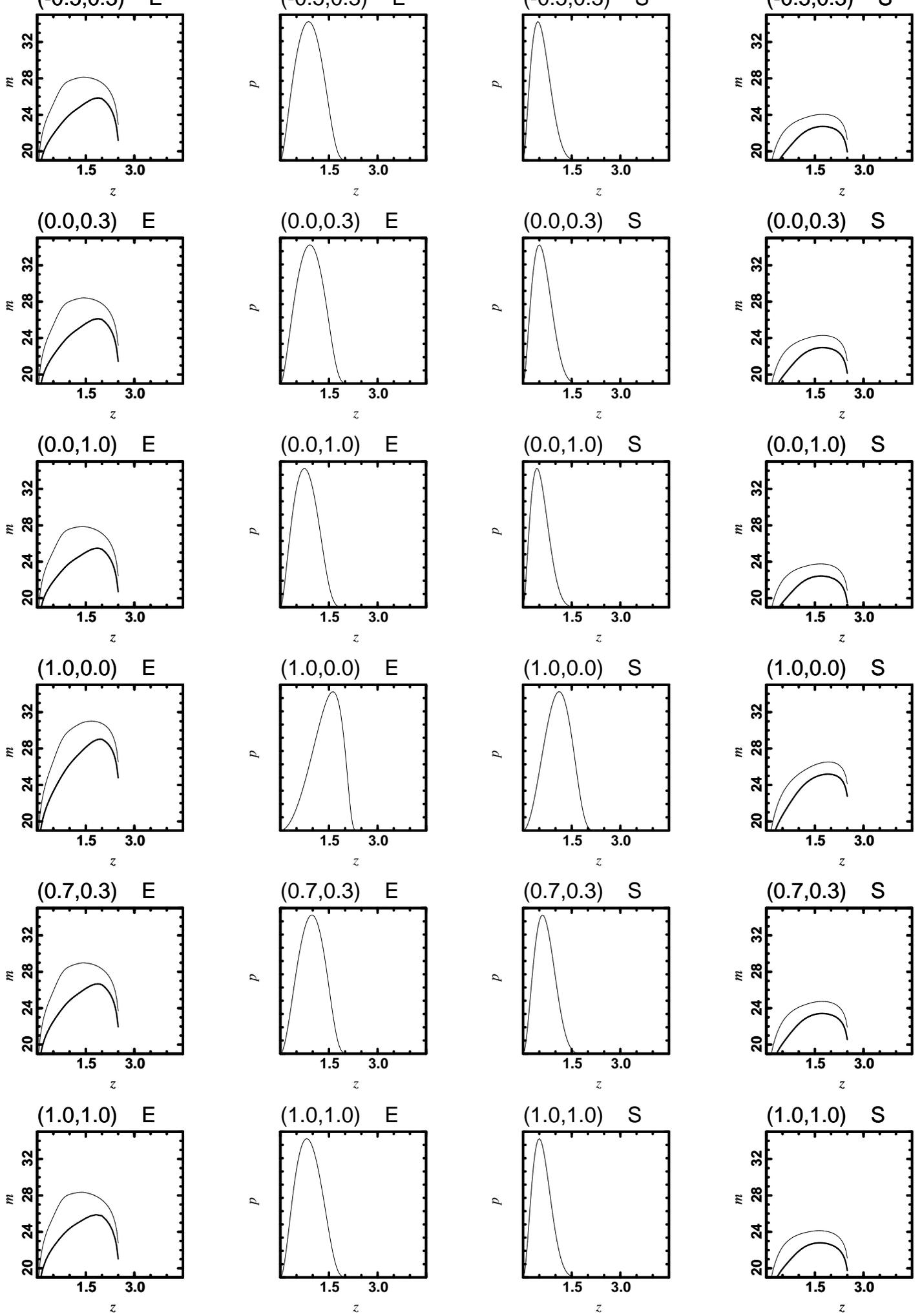

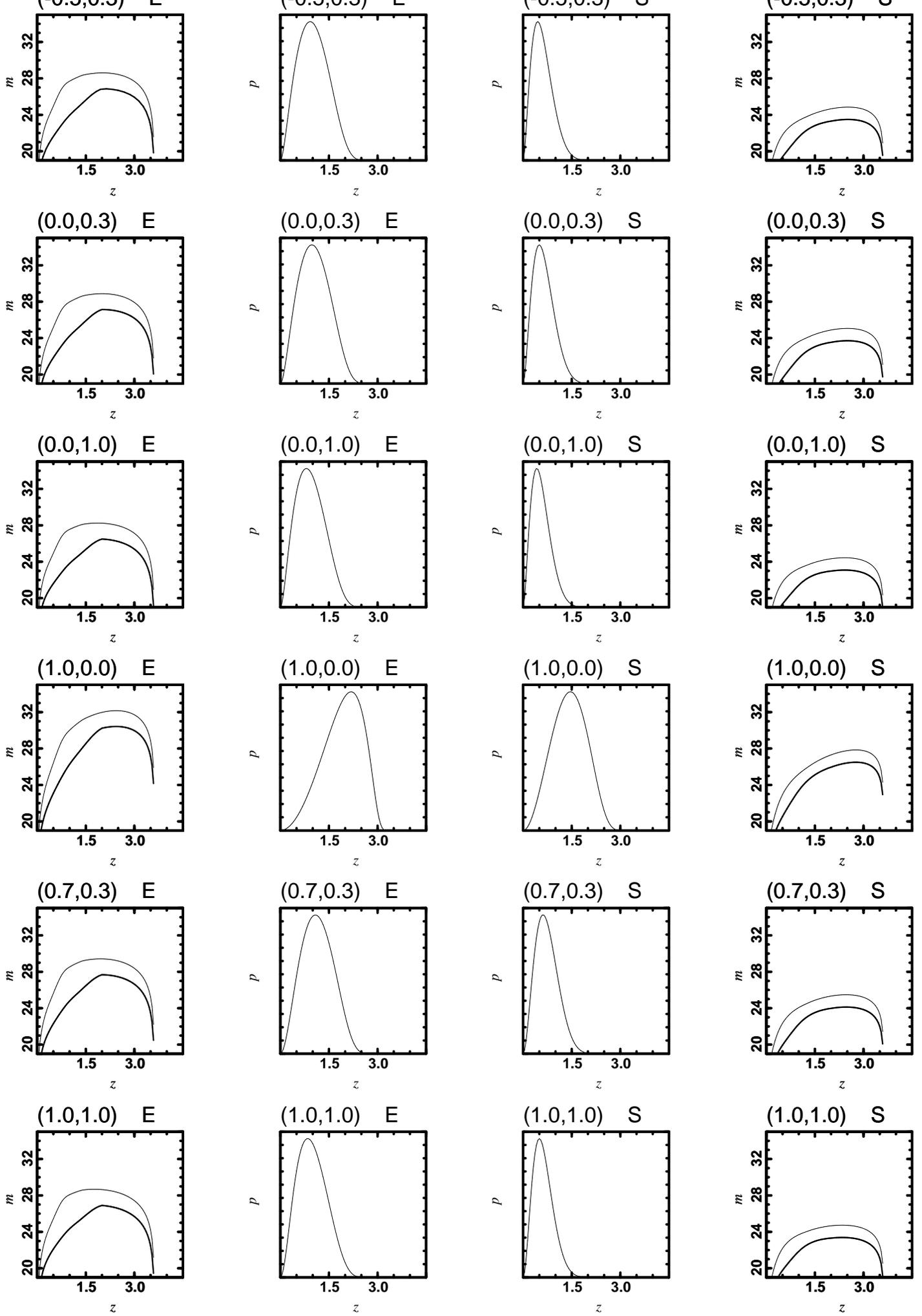

# Predicted lens redshifts and magnitudes for gravitational lenses

## Phillip Helbig

Hamburger Sternwarte, Gojenbergsweg 112
D-21029 Hamburg, Germany

**Abstract**

For suitable gravitational lens systems with unknown lens redshifts, the redshifts and brightnesses (in different colours) of the lenses are predicted for a variety of cosmological models, for both elliptical and spiral galaxy lenses. Besides providing hints as to which systems should be observed with a realistic chance of measuring the lens redshifts, which are needed for detailed lensing statistics and for modelling the lenses, these calculations give a visual impression of the influence of the cosmological model in gravitational lensing.

## a. Introduction

Using basic gravitational lens theory and standard astrophysical approximations, it is possible to calculate, for a given cosmological model and galaxy type, the probability $p$ of finding the lens at a given redshift as well as the brightness of the lens. The observables which have to be known are only the source redshift $z_s$ and the image separation. The basic lens theory is just the lens equation for the singular isothermal sphere, which is a good enough model for lensing statistics (Kraus and White 1992); the needed astrophysical approximations are the Faber-Jackson and Tully-Fisher relations linking the absolute luminosity and the velocity dispersion for elliptical and spiral galaxies, respectively, and the Schechter luminosity function.

The singular isothermal sphere, having an image separation $2a$, where $a$ is the radius of the Einstein ring, *independent* of the relative angular positions of source and lens, allows one to define a cross section for multiply imaging a background source by a single lens, which is just the area ($\approx \pi a^2$) within which the two images will be of comparable brightness (otherwise one image will be too faint to be seen). The relative probability $p$ is proportional to this cross section, the relative numbers of lenses of the appropriate mass and to the volume element $dV/dz$ at the given redshift. The cross section for a single lens is not constant as a function of redshift since the cross section depends on redshift-dependent angular size distances and since the mass ($\rightarrow$ velocity dispersion) needed to produce the observed image separation depends on the redshift. The relative numbers of such galaxies one can get from the Schechter function after converting the velocity dispersion to an absolute magnitude *via* the Faber-Jackson/Tully-Fisher relation. The volume element can be calculated in the standard way.

The cosmological model influences the relative probability through the volume element and through the various angular size distances in the lens equation;

the latter effect influences the needed mass to produce an observed image separation and thus determines the relative number of lenses as well as the cross section for a single lens.

Since the Faber-Jackson/Tully-Fisher relation provides the absolute magnitude, the apparent magnitude, which depends on the redshift and the cosmological model, can be calculated with standard methods. The source redshift $z_\mathrm{s}$ not only of course provides an upper limit to the lens redshift, but also affects the angular size distances between observer and source ($D_\mathrm{s}$) and lens and source ($D_\mathrm{ds}$). The luminosity distance to the lens is related to the angular size distance $D_\mathrm{d}$; this latter distance and the other two angular size distances just mentioned influence $p(z)$.

## b. Theory

I make the 'standard assumptions' that the Universe can be (approximately) described by the Robertson-Walker metric and that lens galaxies can be modelled as non-evolving singular isothermal spheres (SIS). In order to be able to calculate the quantities $p$ and $m$, the relative probability of finding the lens at a given redshift and its apparent magnitude—for a given cosmological model and galaxy type—knowing the source redshift $z_\mathrm{s}$ and image separatation $2a$, one can only examine gravitational lens systems of multiply imaged sources with a small image separation ($\rightarrow$ probably a single galaxy lens with negligible cluster influence) with a measured source redshift.

Making use of the fact that the SIS produces a constant deflection angle, i.e., independent of the position of the source with respect to the optical axis (defined as passing through observer and lens), one can define the angular cross section $\pi a^2$ of a *single* lens for 'strong' lensing events (Turner et al. 1984):

$$\pi a^2 = 16\pi^3 \left(\frac{v}{c}\right)^4 \left(\frac{D_\mathrm{ds}}{D_\mathrm{s}}\right)^2, \tag{1}$$

where $v$ is the one-dimensional velocity dispersion of the lens galaxy, $c$ the speed of light and $D_\mathrm{ds}$ ($D_\mathrm{s}$) the angular size distance between lens and source (observer and source). Following Kochanek (1992), one can arrive at an expression for the optical depth for a given set of observables as follows.

For a given mass distribution, cosmological model, image separation $2a$ and source redshift $z_\mathrm{s}$, $p(z_\mathrm{d})$ is of course proportional to the number of lenses of the mass required to produce the observed image separation per $z_\mathrm{d}$-interval and to the cross section for strong lensing events. ($z_\mathrm{d}$ is the redshift of the lens.) In order to arrive at an expression for $p(z)$ for a *fixed image separation*, one thus needs to know the relative number of lenses which, under the given circumstances, can produce this image separation. This can be calculated by using the Schechter luminosity function (Schechter 1976)

$$\frac{\mathrm{d}n}{\mathrm{d}L} = \frac{n*}{L*}\left(\frac{L}{L*}\right)^\alpha \exp\left(-\frac{L}{L*}\right) \tag{2}$$

as well as the Faber-Jackson and Tully-Fisher relations (Faber & Jackson 1976, Tully & Fisher 1977)

$$\frac{L}{L*} = \left(\frac{v}{v*}\right)^\gamma \tag{3}$$

which give the dependence of the velocity dispersion on the luminosity for elliptical and spiral galaxies, respectively. Bringing in the familiar parameters and dropping all terms which are concerned only with normalisation, one arrives at the expression

$$p(z_\mathrm{d}) = (1+z_\mathrm{d})^2 \frac{a}{a*} \frac{\gamma}{2} \left(\frac{a}{a*} \frac{D_\mathrm{s}}{D_\mathrm{ds}}\right)^{\frac{\gamma}{2}(1+\alpha)} D_\mathrm{d}^2 \frac{1}{\sqrt{Q(z_\mathrm{d})}} \exp\left(-\left(\frac{a}{a*} \frac{D_\mathrm{s}}{D_\mathrm{ds}}\right)^{\frac{\gamma}{2}}\right), \quad (4)$$

where $a* := 4\pi \left(\frac{v*}{c}\right)$ ($v* := v$ of an $L*$ galaxy), $\gamma$ is the Faber-Jackson/Tully-Fisher exponent, $\alpha$ the Schechter exponent, $D_\mathrm{d}$ the angular size distance between the observer and the lens and

$$Q(z_\mathrm{d}) := \Omega_0(1+z_\mathrm{d})^3 - (\Omega_0 + \lambda_0 - 1)(1+z_\mathrm{d})^2 + \lambda_0 \quad (5)$$

Equation (4) is independent of the Hubble constant since the dependences on $H_0$ in the angular size distances and in the Faber-Jackson/Tully-Fisher relation cancel. In order to facilitate comparison the 'standard values' $-1.1$, 2.6, 4, 144 km/s and 276 km/s are used for the Schechter exponent, the Tully-Fisher exponent, the Faber-Jackson exponent, $v*_\mathrm{spiral}$ and $v*_\mathrm{elliptical}$, respectively. (The value for $v*_\mathrm{elliptical}$ includes the factor $(3/2)^{\frac{1}{2}}$ advocated by Turner et al. (1984) and so elliptical galaxies here correspond to the $c = 2$ models examined by Kochanek (1992).)

The optical depth depends on the cosmological model through $Q(z_\mathrm{d})$ as well as through the angular size distances, because of the fact that $D_{ij} = D_{ij}(z_i, z_j; \lambda_0, \Omega_0, \eta)$. The influence of $\eta$, which gives the fraction of homogeneously distributed, as opposed to compact, matter is felt only in the calculation of the angular size distances, whereas the cosmological model in the narrower sense makes its influence felt here as well as through $Q(z_\mathrm{d})$.

In general, there is no analytic expression for the $D_{ij}$; they can be obtained by the solution of a second-order differential equation. (See Kayser (1985) for the derivation of the differential equation, also Linder (1988) for a more general formulation. For an equivalent derivation for $\lambda_0 = 0$ see Schneider et al. (1992). Kayser, Helbig & Schramm (1995) give a general discussion and an easy-to-use numerical implementation.) If one has an efficient method of calculating the angular size distances, it is easy to evaluate Eq. (4) for various world models described by the parameters $\lambda_0$, $\Omega_0$ and $\eta$.

Equation (4) is insensitive to finer points of the mass model such as core radius and ellipticity (cf. Krauss & White (1992), Narayan & Wallington (1992)), basically because the SIS is a good enough model. The fact that Eq. (4) is very nearly independent of $\eta$ is due to this particular combination of angular size distances; in other cases, $\eta$ can have an influence comparable to that of the parameters $\lambda_0$ and $\Omega_0$.

If one is interested in the probability of *finding* the lens, one needs not only $p(z_\mathrm{d})$ but also $m(z_\mathrm{d})$. From the quantities lens redshift $z_\mathrm{d}$, $z_\mathrm{s}$, $a$ and galaxy type one can use Eq. (1) to calculate the velocity dispersion $v$, transform this to an absolute luminosity using the Faber-Jackson or Tully-Fisher relation and then calculate the apparent magnitude as a function of $z_\mathrm{d}$ for the given cosmological model (given by the angular size distance up to powers of $(1+z_\mathrm{d})$ and $K$-corrections). The apparent luminosity of the lens galaxy was calculated in the $B$ (Azusienis & Straizys 1969) and $R$ (Johnson 1965) bands using the $K$-corrections of Coleman, Wu & Weedman (1980) and applying a standard $B - R$ correction for $R$ (since the $B$-band Faber-Jackson and Tully-Fisher relations were used). Since these $K$-corrections are based on displacement of standard spectra at $z = 0$ which extend into the UV-band, they are given only up to $z = 2.0$, where evolutionary effects would in any case have to be considered. (For $z_\mathrm{d} > 2$, a flat spectrum in the relevant interval was assumed, with a value such that the $K$-correction is continuous. Since no constraint was applied to the derivative, the $K$-correction is not completely smooth at this point.) The error introduced by this so defined $K$-correction is small, since for small $z$ evolutionary effects for normal galaxies are small (Kron 1995) and for larger $z$ the apparent magnitude $m$ is determined not so much by the cosmology as by the absolute magnitude $M$, since the required mass—and hence $M$—increases rapidly for large $z$, diverging for $z = z_s$. (Since the *probability* of finding a lens at such a large redshift declines exponentially due to the shape of the Schechter function, one would not in practice expect to observe such large apparent magnitudes.)

For a given cosmological model, image separation $2a$, $z_s$ and galaxy type the apparent magnitude is given by

$$m = M* - 2{,}5\frac{\gamma}{2}\log\left(\frac{\hat{a}}{\hat{a}*}\frac{D_s}{D_{ds}}\right) + 5\log D_L - 5 + K \tag{6}$$

where $M*$ is the absolute magnitude of an $L*$ galaxy, $K$ the $K$-correction and

$$a* = 4\pi\left(\frac{v*}{c}\right)$$

I use $M_{B*} = -19.9 + 5\log h$ (siehe Efstathiou *et al.* 1988) and a $(B - R)$ of 1.8 (1.3) for elliptical (spiral) galaxies (Peletier 1989). The luminosity distance is given by

$$D_L = D_d(1 + z_d)^2 \tag{7}$$

My definition of the luminosity distance and $K$-correction conforms to contemporary standard usage. (See Sandage 1995 for a thorough discussion.) Since the luminosity distance itself and $M*$ are both proportional to $H_0$, the value of the Hubble constant cancels out.

Altogether, the error due to all uncertainties in the calculated magnitudes in the relevant redshift interval—colour scatter between individual galaxies, scatter in the Faber-Jackson/Tully-Fisher relations and the equation of the observed image separation with the critical radius of the lens, which neglects fine points of the mass model—are probably about a magnitude or so (*cf.* Kochanek (1992) where the magnitudes are calculated in a slightly different way.)

## c. Calculations

Since the dependence on $\eta$ is known to be weak, I have chosen to fix $\eta$ at 0.5, which is a value consistent with all cosmological models examined here. To look at the influence of the cosmological model, $(\lambda_0, \Omega_0)$ values of $(-0.5, 0.3)$, $(0.0, 0.3)$, $(0.0, 1.0)$, $(1.0, 0.0)$, $(0.7, 0.3)$ and $(1.0, 1.0)$ were used. These values were chosen to satisfy the majority of the following constraints:

- compatibility with all relatively certain and well-understood observations
- maximisation of the differences due to the cosmological model within the above area
- inclusion of several 'standard models' for purposes of comparison
- limitation of the size of the poster

The cosmological models examined here are thus not meant to be exhaustive but merely illustrative and somewhat representative.

For each gravitational lens system studied, for each cosmological model—described by $(\lambda_0, \Omega_0)$—$p(z_d)$ as well as the lens brightness (in blue (thin curves) and red (thick curves)—presented in the same plot) were calculated. This was done separately for elliptical (E) and spiral (S) galaxies. Details of the gravitational lens systems used are presented in Table 1.

| name | images | $a$ | source | $m_{\text{source}}$ | $z_s$ |
|---|---|---|---|---|---|
| 0952-01 | 2 | 0.45 | QSO | $\Delta$ I = 1.35 | 4.5 |
| 1009-025 | 2 | 0.775 | QSO | R = 17.6<br>R = 20.0 | 2.74 |
| 1104-1805 | 2 | 1.55 | QSO | B = 16.2<br>B = 18.0 | 2.319 |
| 1208+1011 | 2 | 0.225 | QSO | V = 17.5<br>V = 19.0 | 3.803 |
| 1413+117 | 4 | 0.55 | QSO | R = 18.3<br>R = 18.5<br>R = 18.6<br>R = 18.7 | 2.55 |
| 1422+231 | 4 | 0.65 | QSO | R = 16.5 (A–D) | 3.62 |

Table 1. Gravitational lens systems used
For these systems with unknown lens redshifts, the probability of finding the lens at a given redshift and the lens brightness were calculated for a few different cosmological models. (For references see Refsdal & Surdej (1994) and references therein.)

## d. Results and Discussion

Examination of the plots indicates that the probability of finding the lens at a given redshift $p(z_{\text{d}})$ peaks at roughly between one-third and two-thirds of the source redshift for ellipticals, and at noticeably smaller redshifts for spirals. This difference is due principally to the fact that $v*$ for spiral galaxies is substantially lower, so that the required galaxy typically has a velocity dispersion such that the relative number of such galaxies declines more rapidly with increasing $v$ and thus in general increasing $z$ (although the dependence is of course not the same). Also due to this, *if* the lens is a spiral galaxy, then in general it will be brighter than the corresponding elliptical lens, both because it is probably at a lower redshift and because spirals are generally brighter for the same redshift. This might explain why 2 out of 6 lens galaxies with *known* redshifts are spirals, although the fraction of spirals among lens galaxies on the whole is expected to be much smaller. (This is principally because the core radius is not completely negligible; see, *e. g.*, Fukugita *et al.* (1992).) (The differences in colour between spiral and elliptical galaxies as a function of redshift is simply a consequence of the differing spectral energy distribution.)

Except in the case of $1104-1805$, since $m(z_{\text{d}})$ is so steep, selection effects will probably cause those lenses which happen to have a low redshift to be found, regardless of the cosmological model. This means that the probability of *finding* the lens at a given redshift is *not* given by $p(z_{\text{d}})$—this gives the probability of the lens being at a given redshift, whether it can be observed or not.

The de Sitter model (1.0,0.0) is an extreme limiting case (and of course ruled out because $\Omega_0 = 0$ in this model); if one neglects it, then one can make a relatively robust prediction for the redshift and brightness of the lens galaxy in $1104-1805$, since in this case $p(z_\mathrm{d})$ peaks at approximately the same redshift in all cosmological models, the width of the probability distribution is small, and $m(z_\mathrm{d})$ is comparatively not very steep. (All of these effects are a consequence of the relatively large image separation in this system, which is also larger than that in any of the comparable systems with *known* redshifts. This system also has the smallest source redshift of the systems with unknown lens redshifts (though with one exception larger than all source redshifts of systems with known lens redshifts) which also contributes somewhat to the effect. Of course, a contribution by an unseen cluster would invalidate the approximations used here.) Roughly, the lens should lie in the range $0.3 < z_\mathrm{d} < 0.7$ and be brighter than about 21.5 in $R$, which means that it could be detected (or that we live in a cosmological model nearer the de Sitter model). If there is not a strong selection effect in favour of spirals over ellipticals—and there isn't in this case because even an elliptical galaxy should be bright enough to be detected at or near the most probable redshift—then one would expect the lens to be an elliptical. This is probably the case, and a spiral lens would be so bright that it probably would have been found already. The fact that no lens has yet been found in this system can have one of three reasons: our cosmological model is near the de Sitter model, there is an unseen cluster responsible for the large image separation and thus the approximations used here break down, or the brightness of the images—brighter than comparable systems with of without known lens redshifts—makes measuring the lens redshift difficult. (Of course, any combination of these could also be the case.)